\documentclass[aps,prb,reprint,amsmath,longbibliography,twocolumn]{revtex4-2}
\usepackage{graphicx}
\usepackage{enumerate}
\usepackage{refcount}
\usepackage{fmtcount}
\usepackage{amssymb}
\usepackage{tabularx}
\usepackage{float}
\usepackage{hyperref}
\usepackage[usenames,dvipsnames]{color}
\usepackage[normalem]{ulem}
\usepackage{bm}

\hypersetup{pdfstartview=FitH,pdfpagemode=UseOutlines,colorlinks,
citecolor=blue,linkcolor=blue,urlcolor=blue}

\definecolor{blue(munsell)}{rgb}{0.0, 0.5, 0.69}

\begin{document}


\title{Pressure-tuned many-body phases through ${\bf \Gamma}$-K valleytronics in moir\'e bilayer WSe$_2$}

\author{Marta Brzezi\'nska}
\affiliation{Institute of Physics, Ecole Polytechnique Fédérale de Lausanne (EPFL), CH-1015 Lausanne, Switzerland}
\author{Sergii Grytsiuk}
\affiliation{Institute for Molecules and Materials, Radboud University, NL-6525 AJ Nijmegen, The Netherlands}
\author{Malte R\"osner}
\affiliation{Institute for Molecules and Materials, Radboud University, NL-6525 AJ Nijmegen, The Netherlands}
\author{Marco Gibertini}
\affiliation{Dipartimento di Scienze Fisiche, Informatiche e Matematiche, University of Modena and Reggio Emilia, IT-41125 Modena, Italy}
\affiliation{Centro S3, CNR-NANO Istituto Nanoscienze, IT-41125, Modena, Italy}
\author{Louk Rademaker}
\affiliation{Department of Quantum Matter Physics, University of Geneva, CH-1211 Geneva, Switzerland}

\date{\today}

\begin{abstract}
Recent experiments in twisted bilayer transition-metal dichalcogenides have revealed a variety of strongly correlated phenomena. 
To theoretically explore their origin, we combine here ab initio calculations with correlated model approaches to describe and study many-body effects in twisted bilayer WSe$_2$ under pressure.
We find that the interlayer distance is a key factor for the electronic structure, as it tunes the relative energetic positions between the K and the $\Gamma$ valleys of the valence band maximum of the untwisted bilayer.
As a result, applying uniaxial pressure to a twisted bilayer induces a charge-transfer from the K valley to the flat bands in the $\Gamma$ valley.
Upon Wannierizing moiré bands from both valleys, we establish the relevant tight-binding model parameters and calculate the effective interaction strengths using the constrained random phase approximation. With this, we approximate the interacting pressure-doping phase diagram of WSe$_2$ moiré bilayers using self-consistent mean field theory.
Our results establish twisted bilayer WSe$_2$ as a platform that allows the direct pressure-tuning of different correlated phases, ranging from Mott insulators, charge-valley-transfer insulators to Kondo lattice-like systems.
\end{abstract}

\maketitle


\section{Introduction}

Moir\'e semiconductor heterostructures have proven to be an ideal platform for creating and manipulating nontrivial, correlated electron phases~\cite{Mak.2022}. 
The observed phenomena range from Mott criticality~\cite{Ghiotto.2021,Li.202109b}, to Wigner-Mott crystals~\cite{Regan.2020, Jin.2021, Tang.2023}, exciton condensation~\cite{Wang.2019wdv} and the quantum anomalous Hall effect~\cite{Li.2021nyj}.
The main constituents in these experiments are moir\'e bilayers of transition-metal dichalcogenides (TMDs), which come in two forms: heterobilayers, where the two layers are different materials, such as WSe$_2$/WS$_2$; and homobilayers, where the two layers are the same, such as twisted bilayer WSe$_2$.

Whereas twisted bilayer WSe$_2$ was one of the first materials to reveal correlated insulator physics in transport experiments~\cite{Wang.2020}, the correct low-energy description of the relevant flat bands remains a subject of debate. The two main candidates are a Kane-Mele topological insulator model~\cite{Wu.2019,Devakul.2021yq2s} or a triangular Hubbard model~\cite{Pan.202045,Zang.2021,Ryee.2023,Tscheppe.2024}, both derived from states at the K point in the Brillouin zone. The difference between these two flat band models is rooted in the precise symmetries of the moir\'e potential, which is difficult to estimate from first principles as scanning tunneling microscopy results seem in contradiction with the prevailing density functional theory (DFT) predictions~\cite{Zhang.2020}.
At the same time, photoemission spectroscopy~\cite{Pei.2022,Gatti.2023} was not able to resolve the moir\'e potential and flat bands at the K point, but did observe large moir\'e gaps and dispersionless bands emerging around the $\Gamma$ point. Ideally, if one could access the flat bands at the $\Gamma$ point, it would be possible to realize much stronger electron correlations~\cite{Zhang.2021,Angeli.2021}.

In this manuscript, we show how to correctly align predictions from density functional theory with the experimental observations on twisted bilayer WSe$_2$. Furthermore, we show that uniaxial pressure can be used to tune the relative energy (and thus occupation) of $\Gamma$ and K valley states. This, in turn, allows for a plethora of correlated phases: antiferromagnetic insulators and valley-charge-transfer insulators, and their doped counterparts including a Kondo lattice regime. This renders bilayer WSe$_2$ under uniaxial pressure an ideal platform to study correlated metallic states.

In Sec.~\ref{Sec:AbInitio}, we use DFT to calculate properties of six differently stacked untwisted bilayer WSe$_2$. These results are used in Sec.~\ref{Sec:Moire} to derive the effective flat band physics for both parallel and antiparallel twisted bilayer WSe$_2$. Using estimates for the relevant Coulomb interaction matrix elements based on a Wannierization of the flat bands, we derive a full many-body model and discuss its interacting phase diagram in Sec.~\ref{Sec:PhaseDiagram}, revealing distinct correlated and charge-transfer insulators.

\section{Ab-initio results}
\label{Sec:AbInitio}

\begin{figure*}
    \centering
    \includegraphics[width=\textwidth]{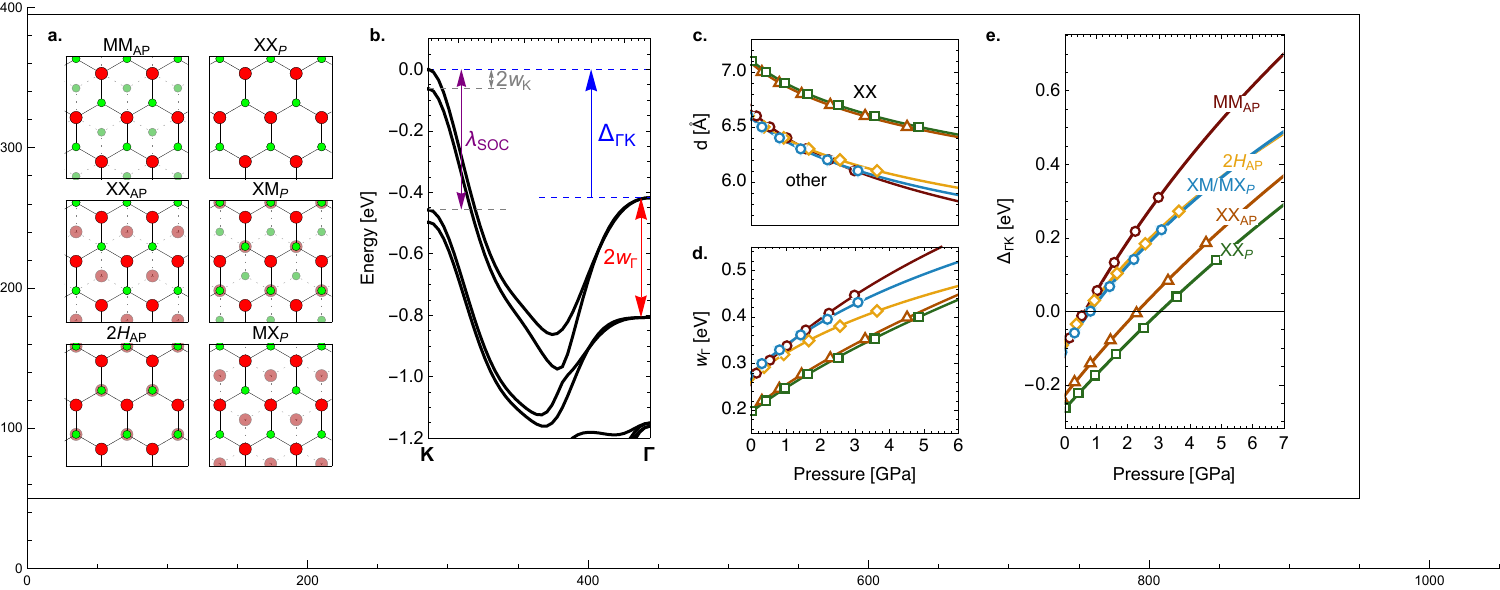}
    \caption{{\bf a.} Top view of the six possible stacking of untwisted bilayer WSe$_2$. Red corresponds to the W atoms, green to the Se atoms; bright colors are the top layer, faded colors the bottom layer. The three parallel (P) stackings are obtained by starting with the XX stacking and performing a lateral shift of one layer; the three antiparallel (AP) stackings are obtained from the parallel ones by rotating one layer by 180 degrees.
    {\bf b. } The electronic structure of untwisted bilayer WSe$_2$ is characterized by the valence band maxima at $\Gamma$ and K. Here we show an illustrative band structure for the XX$_{\rm P}$ stacking, indicating the valley offset $\Delta_{\Gamma {\rm K}}$, the interlayer splittings $w_{\Gamma/K}$ and the spin-orbit coupling $\lambda_{\rm SOC}$.
    {\bf c.} The interlayer distance $d$, defined as the W-W distance along the $z$-direction, is the dominant physical parameter that determines the electronic structure and can be tuned by applying uniaxial pressure. Here we show the interlayer distance as a function of applied pressure.
    {\bf d.} A decrease in interlayer distance leads to an increase in interlayer hopping $w$ at the $\Gamma$-point between W $d_{z^2}$ orbitals. 
    {\bf e.} Finally, the application of pressure strongly influences the valley offset $\Delta_{\Gamma {\rm K}}$. Even though the precise value is hard to predict, here we show the tendency that for different stackings, the $\Delta_{\Gamma {\rm K}}$ increases with increasing pressure and thus with decreasing interlayer distance $d$.}
    \label{fig:abinitio}
\end{figure*}

Monolayer WSe$_2$ has a honeycomb structure with the metal W on one sublattice and two Se on the other sublattice site, vertically displaced in opposite directions with respect to the plane of W atoms. {\em Untwisted} bilayer WSe$_2$ (bWSe$_2$) can therefore be realized in six different high-symmetry stackings, which are visualized in Fig.~\ref{fig:abinitio}a. We performed first-principles calculations using \textsc{Quantum ESPRESSO}~\cite{Giannozzi.2009, Giannozzi.2017,Giannozzi.2020} to compute the electronic properties of these differently stacked bWSe$_2$, and how their properties change under applied pressure~\cite{SupplInfo}.

Fig.~\ref{fig:abinitio}b shows an illustrative band structure of bWSe$_2$. Typically, the valence band maximum is located at K, where the orbital content is predominantly $d_{xy} \pm i d_{x^2 - y^2}$ on the W atoms~\cite{Liu.2013}. The bands have a large spin-orbit splitting $\lambda_{\rm SOC}$ and a small interlayer hybridization $w_{\rm K}$. Additionally, there exists a local valence band maximum at $\Gamma$. Here, the orbital content is predominantly $d_{z^2}$, the bands are spin-degenerate but layer-hybridized with interlayer hopping $w_\Gamma$. The energy difference between the $\Gamma$ and K valley is characterized by the valley offset $\Delta_{\Gamma {\rm K}}$. In photoemission spectroscopy, $\Delta_{\Gamma {\rm K}}$ is observed to be about $-90$ meV, meaning the K valley is higher in energy~\cite{Gatti.2023}.

We found that the electronic parameters (energies and splittings at K and $\Gamma$) are extremely sensitive to one specific structural parameter: {\em the interlayer distance} $d$, defined as the separation between W planes along the $z$-axis. Other structural parameters, such as the precise locations of the Se atoms and the lattice constant, are subleading~\cite{SupplInfo}. Unfortunately, layered materials such as bWSe$_2$ cannot be correctly described by only (semi)-local functionals, because they do not capture the long-range nature of the interlayer Coulomb and van der Waals (vdW) interactions. An accurate estimation of the interlayer distance $d$ from first principles is thus challenging, as it requires either to consider vdW-compliant functional extensions~\cite{Klimes.2012} or to resort to more sophisticated --but computationally expensive-- approaches such as the Random Phase Approximation (RPA)~\cite{Lu.2009} or Many-Body dispersion (MBD)~\cite{Tkatchenko.2012} methods.

\begin{figure*}
    \centering
    \includegraphics[width=\textwidth]{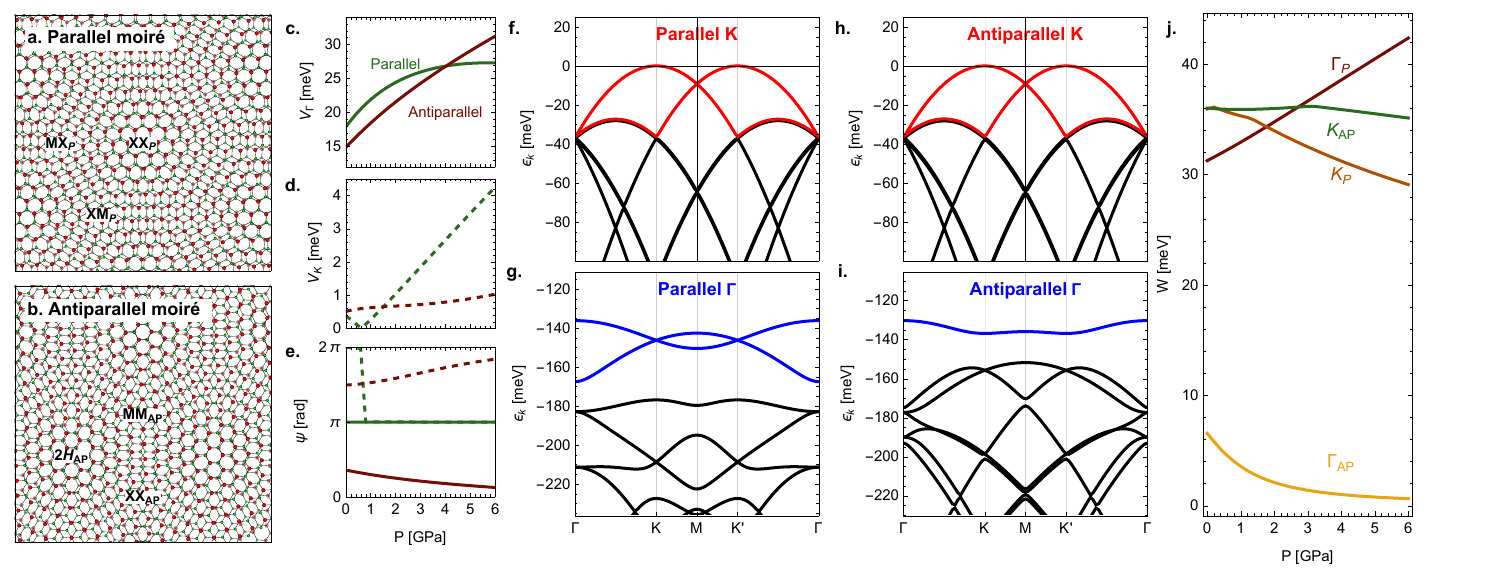}
    \caption{{\bf a.} The parallel moir\'e structure, obtained from untwisted XX$_{\rm P}$ stacking and applying a small twist angle, consists of a hexagonal structure with locally XX$_{\rm P}$, XM$_{\rm P}$ and MX$_{\rm P}$ stacking. Note that the XM$_{\rm P}$ and MX$_{\rm P}$ regions are symmetry-related.
    {\bf b.} The antiparallel moir\'e structure consists of a hexagonal structure with locally MM$_{\rm AP}$, 2H$_{\rm AP}$ and XX$_{\rm AP}$ stacking. There is no symmetry relation between the different regions of the moir\'e unit cell.
    {\bf c/d/e.} By tracing the local valence band maxima at $\Gamma$ or K at the different stackings in the moir\'e unit cell, we can extract the moir\'e potential following Eqs.~\eqref{eq:moire_potential} and \eqref{eq:moire_potential2}. Here we show the dependence of the moir\'e potential amplitude at ${\Gamma}$ (c) and K (d) and the moir\'e potential phase (e) on applied uniaxial pressure. Note that the moir\'e potential at $\Gamma$ is an order of magnitude larger than at K.
    {\bf f/g.} The electronic band structure of parallel moir\'e structures at zero uniaxial pressure, throughout the mini-Brillouin zone, for a twist angle $\theta = 3^{\circ}$. The band structure at K (f) is barely affected by the moir\'e potential and does not lead to the opening up of significant moir\'e gaps. On the other hand, there appear isolated moir\'e flat bands at ${\Gamma}$ (g) with honeycomb symmetry, consistent with ARPES observations~\cite{Gatti.2023}.
    {\bf h/i.} The electronic band structure of antiparallel moir\'e structures at zero uniaxial pressure, throughout the mini-Brillouin zone, for a twist angle $\theta = 3^{\circ}$. Similar to the case of parallel moir\'e structures, the band structure at K (h) shows no signs of moir\'e gaps. The isolated subset of moir\'e flat bands at ${\Gamma}$ (i) has a triangular lattice symmetry.
    {\bf j.} The bandwidth of moir\'e flat bands at the top of the $\Gamma$ and K valley in parallel and antiparallel moir\'e structures. Most notably are the $\Gamma$-states in the antiparallel system, which attain vanishingly small bandwidth under pressure.}
    \label{fig:moire}
\end{figure*}

In the specific case of bWSe$_2$, we find that determining an accurate value for $d$ is particularly demanding, with an enormous range in predicted interlayer distances, ranging from $d=6.462$~\AA{} to $d=7.800$~\AA{}, obtained  by relaxing the atomic positions of the XM$_{\rm P}$ stacking for different pseudopotentials, functionals and vdW corrections (see \cite{SupplInfo}, Sec. IB for details). 
This, in turn, leads to a range of valley offsets from $\Delta_{\Gamma {\rm K}} = -89.4$ meV to $-475.3$ meV. The tendency, however, is clear: when the interlayer distance is small, the energy of the $\Gamma$ valley is higher. 
Physically, this effect stems from the fact that the interlayer hybridization $w_\Gamma$ of the $d_{z^2}$ orbitals is increased when the interlayer distance is decreased. In contrast, the interlayer hoppings involving $d_{xy}/d_{x^2-y^2}$ orbitals are less affected by the change in interlayer distance (and completely vanish in antiparallel stackings due to spin symmetry).
Similarly, we studied how the electronic structure is affected by encapsulating or capping the WSe$_2$ bilayers. Again we found that the interlayer distance $d$ is the most significant determinant for $\Delta_{\Gamma {\rm K}}$, regardless of the environment (see \cite{SupplInfo}, Sec. IC for details).

Although the qualitative tendency under a change of interlayer distance is clear, it does not allow for a quantitative accuracy on energy scales less than roughly 10 meV. For example, a tiny shift in atomic positions of merely 0.01 \AA{} leads to several meV changes in the band energies. 
Therefore, when we derive moir\'e flat bands in the next Section, we cannot provide exact estimates for the moir\'e potential. Rather, we describe the tendencies and infer from experiments (such as ARPES~\cite{Gatti.2023} or STM~\cite{Zhang.2020}) the correct DFT predictions. For definiteness, to incorporate vdW corrections in DFT calculations, we employ the rVV10 functional~\cite{VV10,Sabatini.2013}, known to reproduce the correct RPA trends for the binding energy of layered materials~\cite{Bjorkman.2012}, and that in the specific case of bWSe$_2$ provides lattice parameters and interlayer distances~\cite{SupplInfo} in good agreement with reported RPA results~\cite{He.2014ahh}.

Since the interlayer distance $d$ is the dominant parameter, the most straightforward way to tune the electronic structure is through the application of {\em uniaxial pressure}. To estimate the interlayer distance as a function of pressure, we fixed various W-W distances in the six high-symmetry bilayer stackings, and relaxed the in-plane lattice constant and other atomic positions. The pressure is extracted from ab initio calculations using the derivatives of the energy. The resulting distance-pressure curve is shown in Fig.~\ref{fig:abinitio}c, and is consistent with the Murnaghan relation $P = A (e^{-B (1 - d/d_0)}-1)$ with $A,B$ being stacking-dependent parameters~\cite{Carr.2018,Morales-Durán.2023m1}. 

As the interlayer distance decreases with applied pressure, the interlayer hopping between $d_{z^2}$ orbitals increases, as shown in Fig.~\ref{fig:abinitio}d. This, in turn, causes the valence band maximum to shift from the K point to the ${\Gamma}$ point, see Fig.~\ref{fig:abinitio}e. Depending on the stacking, the valley transition point happens between $P=0.5$ -- $3$ GPa.


\section{Electronic structure in moir\'e bilayers}
\label{Sec:Moire}

Now that we have analyzed the electronic structure of {\em untwisted} bWSe$_2$, we will discuss {\em twisted} bWSe$_2$. This structure exists in two inequivalent forms: so-called parallel or antiparallel stacking, as shown in Fig.~\ref{fig:moire}a,b. The size of the moir\'e unit cell depends on the twist angle $\theta$, given in the continuum limit by $a_M = a_{\rm WSe_2} / \sin \theta$. 
To make a connection between the twisted moiré structures and the untwisted results from previous section, we perform a set of structural relaxations of twisted bilayers using LAMMPS~\cite{Thompson.2022,SupplInfo}.
We look at the angle range $\theta \sim 3-6^{\circ}$ where effects related to lattice reconstruction (such as domain wall network formation) can be neglected~\cite{Weston.2020,Enaldiev.2020,Nielsen.2023}, and which is relevant for the reported transport measurements~\cite{Wang.2020}.
Specifically, we consider parallel and antiparallel supercells at angles $\theta = 3.2, 4.4, 5.1, 6.0$ degrees. 
We quantify the interlayer spacings and bond lengths as a function of the position in the moir\'e unit cell.
Locally, there exists high-symmetry stackings, as shown in Fig.~\ref{fig:moire}a,b. 
Interestingly, we found that there is no significant dependence of the in-plane bond lengths $a$ and interlayer distance $d$ on the twist angle $\theta$. 
Even though the interlayer distance can vary up to 0.3 \AA{} throughout one moir\'e unit cell, it follows the predicted interlayer distances from the untwisted bilayers: the regions with the largest $d$ correspond to the local XX stacking orders. These results are consistent with Ref.~\cite{Vitale.2021}, further details are presented in the Supplementary Information~\cite{SupplInfo}.

Consequently, we can use the ab initio results of the untwisted bilayers of Sec.~\ref{Sec:AbInitio} to predict the electronic structure of twisted bilayers. We follow Refs.~\cite{Wu.2018,Wu.2019}: the monolayer states at $\Gamma$ and K are approximated by a parabolic dispersion, and the effect of the twist is captured by the {\em moir\'e potential} $\Delta({\bf r})$, which is calculated as follows. 
Starting with {\em untwisted} XX$_{\rm P}$ or MM$_{\rm AP}$ stacking, the other high-symmetry stacks are obtained by performing $s =0,1,2$ shifts of the top layer along the vector $\mathbf{r}_0 = ( \mathbf{a}_1 + \mathbf{a}_2) / 3$, where $ \mathbf{a}_1 = a (1, 0)$ and $ \mathbf{a}_2 = a (1/2, \sqrt{3}/2)$. This provides us the sequence of high-symmetry stackings XX$_{\rm P}$ $\rightarrow$ MX$_{\rm P}$ $\rightarrow$ XM$_{\rm P}$ (or MM$_{\rm AP}$ $\rightarrow$ XX$_{\rm AP}$ $\rightarrow$ 2H$_{\rm AP}$). In a {\em twisted} bilayer, the local stacking configuration varies smoothly throughout the moiré unit cell, following the same sequence of stackings. Therefore, the holes in the moiré valence band experience a periodic potential $\Delta (\mathbf{r})$, consistent with the varying energy of the valence band maxima of the untwisted high symmetry stacking. We expand the moiré potential in lowest order of the Fourier expansion over the nearest moiré reciprocal lattice vectors $\mathbf{g}_j$,  $| \mathbf{g}_j | = 2 \pi / a$,
\begin{equation}
\Delta (\mathbf{r}) = \sum_{j = 1}^6 V_j \exp (i \mathbf{g}_j \cdot \mathbf{r})
\label{eq:moire_potential}
\end{equation}
As $\Delta (\mathbf{r})$ is a real threefold-symmetric function, the values of $V_j$ are constrained, 
$V_1 = V_3 = V_5, V_2 = V_4 = V_6, V_1 = V_4^* \equiv V e^{i \psi}$,
where $V$ is the real amplitude and $\psi$ is the phase. The values of $V$ and $\psi$ are found by solving the relation 
\begin{equation}
E_{(\Gamma, {\rm  K}), s}  = V_0 + 6 V \cos \left( \psi + s \cdot  \frac{2\pi}{3} \right),
\label{eq:moire_potential2}
\end{equation}
for $V_0, V, \psi$, where $E_{(\Gamma, {\rm K}), s} $ is the valence band energy at $\Gamma$ or K given the shift $s$. We calculated the values (amplitude and phase $\psi$) of the moir\'e potential for both stackings (parallel/antiparallel) and in both valleys ($\Gamma$/K) as a function of uniaxial pressure, for a fixed twist angle value $\theta = 3^\circ$.

\begin{figure}
    \centering
    \includegraphics[width=\columnwidth]{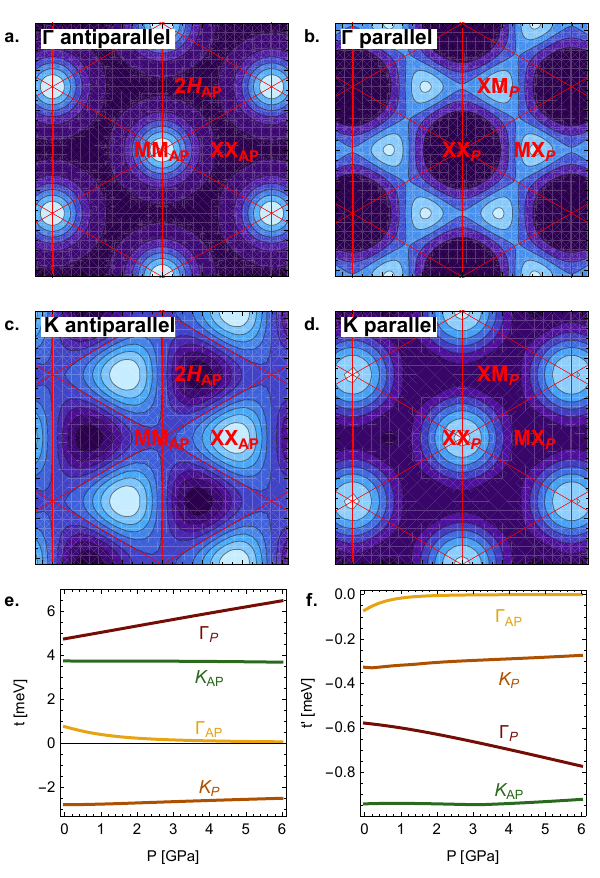}
    \caption{{\bf a.} Visualization of the Bloch wavefunction in the $\Gamma$ valley for antiparallel twisted bilayers, shown as the absolute value squared summed over both layers of the wavefunction. Model parameters are $\theta = 3^\circ$ and $P = 0$ GPa. The electronic states is localized in the MM$_{\rm AP}$ region of the moir\'e unit cell. 
    {\bf b.} Same as a, but for parallel twisted bilayers. We clearly see the emergence of a honeycomb lattice on the XM$_{\rm P}$/MX$_{\rm P}$ positions.
    {\bf c.} Same as a, but for the K valley states in an antiparallel twisted bilayer. The states are not as localized as in the $\Gamma$ valley, and are centered around the XX$_{\rm AP}$ position.
    {\bf d.} Same as c, but for the parallel twisted bilayers. There is less localization compared to the $\Gamma$ valley, and the states are centered at XX$_{\rm AP}$.
    {\bf e/f.} The nearest-neighbor $t$ and next-nearest-neighbor $t'$ tight-binding hopping parameters at $\theta = 3^\circ$ as a function of pressure, for the triangular lattice models relevant for antiparallel twisted bilayers and the K valley of parallel bilayers; as well as for the honeycomb lattice model relevant for the parallel bilayer $\Gamma$ valley states.}
    \label{fig:tightbinding}
\end{figure}

The calculated moir\'e potentials shown in Fig.~\ref{fig:moire}c-e provide two main results. Firstly, the amplitude of the moir\'e potential $V$ is an {\em order of magnitude larger} at $\Gamma$ than at K. Secondly, pressure increases the potential amplitude $V$. Let us now discuss the details of the moir\'e electronic structure, for both parallel and antiparallel structures.

\begin{figure*}
    \centering
    \includegraphics[width=\textwidth]{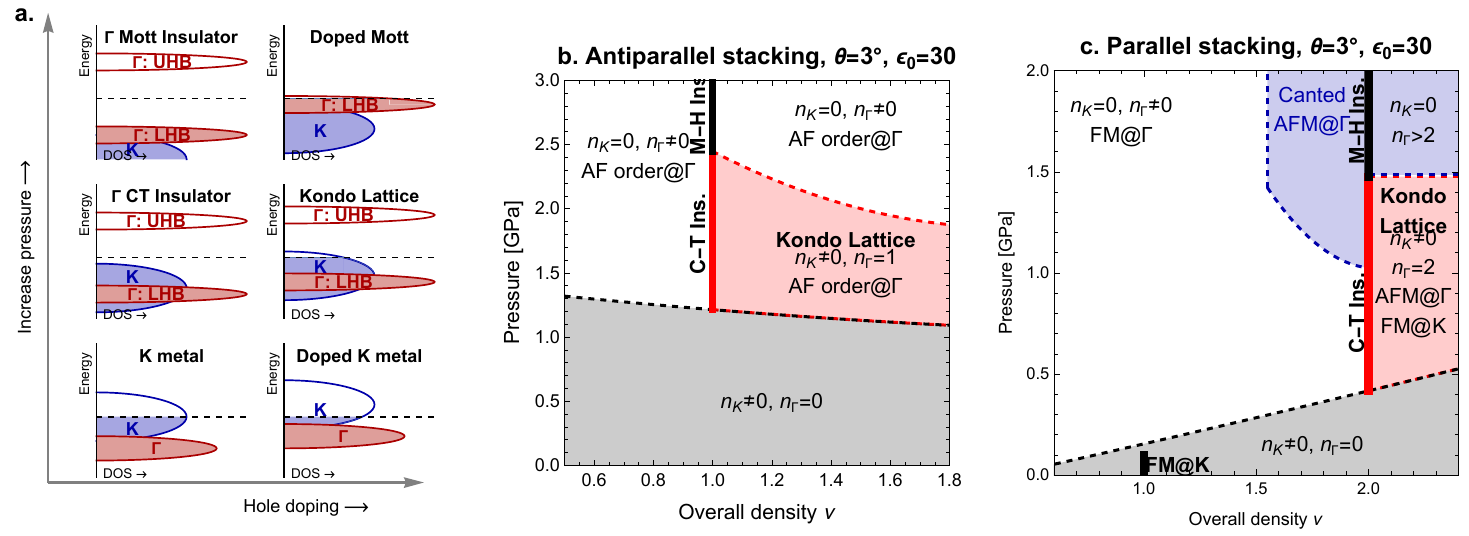}
    \caption{
    {\bf a.} Qualitatively, we can already identify various many-body phases that are expected in tbWSe$_2$ under pressure. Here, we visualize six possible phases by plotting a cartoon density of states versus energy.
    At zero or low pressure, the charge carriers will fill up states in the weakly correlated K valley (bottom row), leading to metallic behavior.
    At half-filling (left column), an increase in pressure leads to a charge-transfer to the strongly correlated $\Gamma$ valley. The $\Gamma$ states will split into a lower and upper Hubbard band (LHB/UHB), yielding an insulator state. Depending on the position of the remaining K valley states, this is a charge-transfer (CT) insulator at intermediate pressures, or a Mott-Hubbard insulator at high pressures. 
    Increasing the hole doping beyond half-filling (right column) now yields either a Kondo lattice model at intermediate pressures, or a doped Mott insulator at high pressures
    {\bf b/c.} The qualitative picture is confirmed by numerical mean field theory calculations. Here we show the resulting phase diagram for antiparallel (b) and parallel (c) stacking at twist angle $\theta = 3^\circ$ and dielectric constant $\epsilon_0 = 30$, as a function of hole density $\nu$ and pressure $P$. 
    We confirm that at a critical pressure the electronic charge is fully shifted from the K to the $\Gamma$ valley. At half-filling of the $\Gamma$ valley states ($\nu =1$ for antiparallel stacking, $\nu = 2$ for parallel stacking), applying pressure induces a transition from a conducting K state to an antiferromagnetic $\Gamma$ charge-transfer insulator (C-T Ins.), followed by a transition to a Mott-Hubbard insulator (M-H Ins.). Increasing the hole density $\nu$ in the charge-transfer insulator leads to a Kondo lattice regime, with localized electrons in the $\Gamma$ valley and conduction electrons in the K valley.}
    \label{fig:phasediagram}
\end{figure*}

In parallel structures, the XM$_{\rm P}$ and MX$_{\rm P}$ are symmetry-related through a $2\pi /3$ rotation followed by a layer-flip. Consequently, the moir\'e potential phase is either $\psi = 0$ or $\pi$. The corresponding moir\'e flat bands at $\Gamma$ (using the method of \cite{Angeli.2021,Zhang.2021}) therefore display a honeycomb symmetry, see Fig.~\ref{fig:moire}g. There is a clear gap between the moir\'e flat bands and the other bands at $\Gamma$. By contrast, the moir\'e potential in the K-valley is very small, leading to the absence of a moir\'e gap opening, see Fig.~\ref{fig:moire}f (at K, we follow the model of Ref.~\cite{Wu.2019}).
In real space, the honeycomb structure at $\Gamma$ is clearly visible in Fig.~\ref{fig:tightbinding}b, with localized states at the XM$_{\rm P}$/MX$_{\rm P}$ points of the moir\'e unit cell. By contrast, the states in the K valley are centered around XX$_{\rm P}$, see Fig.~\ref{fig:tightbinding}d. Note that this is consistent with experimental STM measurements~\cite{Zhang.2020}. The small value of the moir\'e potential and the real-space structure of K valley states we found is consistent with the theoretical work of Ref.~\cite{Pan.202045}.

A similar pronounced difference between K and $\Gamma$ valley states is present in antiparallel twisted bilayers. Due to the lack of symmetry between the different local stackings, now there is no constraint on the moir\'e potential phase. As a result, both $\Gamma$ and K valley states form a triangular lattice in real space, as shown in Fig.~\ref{fig:tightbinding}a,c. Because the moir\'e potential is an order of magnitude larger in the $\Gamma$ valley, only there we find a well-separated moir\'e flat band that is spin-degenerate. However, note that due to the absence of interlayer hopping for K valley states in the antiparallel bilayer, the resulting moir\'e bands are layer degenerate with a spin-layer-valley coupling. 

It is interesting to note that the $\Gamma$-valley states are, for both parallel and antiparallel stackings, spin-degenerate. On the other hand, due to the spin-valley locking in monolayer WSe$_2$, the K-valley states are also degenerate if we take into account both K and K' valleys.

Our calculation of the electronic structure in twisted bilayers allows us to compute the effective tight-binding hopping parameters for upmost moiré bands from the two valleys and two stackings.
The resulting $t$ and $t'$ as a function of pressure are shown in Fig.~\ref{fig:tightbinding}e,f. 
With the exception of the $\Gamma$-valley states in parallel stacking, the hopping parameters $t,t'$ decrease with increasing unixial pressure due to the increasing strength of the moir\'e potential.
This exception at $\Gamma_{\rm P}$ is due to the fact that as the moir\'e potential increases, the moir\'e gap at the mini-Brillouin zone K-points increases faster than at $\Gamma$, and therefore slightly increases the flat band bandwidth. 

Finally, we do not find parameters that put the moir\'e flat bands in the topological regime, as was originally proposed for twisted bilayer MoTe$_2$~\cite{Wu.2019}. All effective band structures discussed here are topologically trivial.

\section{Interacting phase diagram}
\label{Sec:PhaseDiagram}

\begin{figure}
    \centering
    \includegraphics[width=\columnwidth]{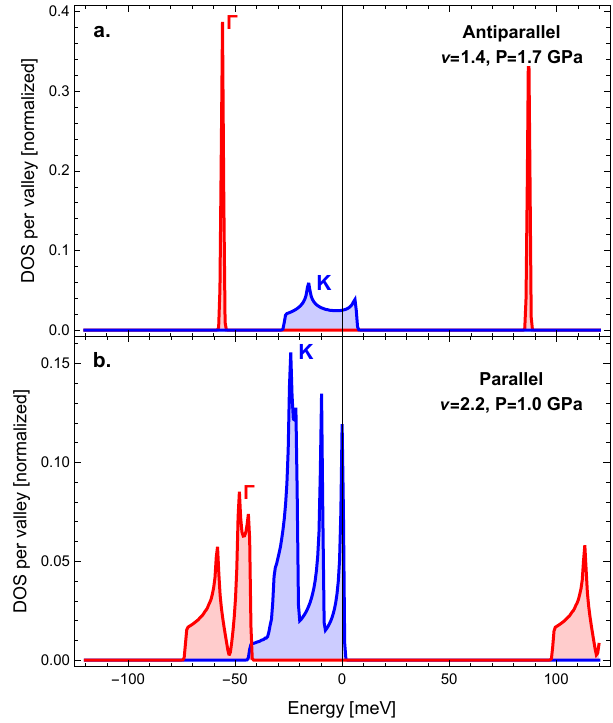}
    \caption{The density of states in the Kondo lattice regime of the phase diagrams of Fig.~\ref{fig:phasediagram}, for antiparallel ({\bf a}) and parallel ({\bf b}) stacking. In both cases the strong correlations of the $\Gamma$ states at half-filling lead to the formation of lower and upper Hubbard bands. Upon doping away from half-filling, the Fermi level ($E=0$) crosses the K valley states that are lying within the gap of the $\Gamma$ states. The result is a system with localized electrons in the $\Gamma$ valley and conduction electrons in the K valley.}
    \label{fig:mft_dos}
\end{figure}

Up till now we have only discussed band structure effects through the application of pressure in twisted bilayer WSe$_2$. The most striking feature was that the 'flattest' bands exist in the ${\Gamma}$ valley, which can be brought to the top of the valence band through pressure. The flatness of these bands prompts us to investigate the effect of interactions. The starting point for this discussion is the Hubbard model of the form
\begin{equation}
    H^{vs} = \sum_{{\bf k } \sigma} \epsilon^{vs}_{{\bf k} \sigma} n_{{\bf k}\sigma} + U^{vs} \sum_i n_{i \uparrow} n_{i \downarrow}
\end{equation}
where $\epsilon^{vs}_{{\bf k} \sigma}$ is the non-interacting dispersion in valley $v = \Gamma$, K with stacking $s = {\rm P}, {\rm AP}$. The sites $i$ run over the honeycomb lattice sites at XM$_{\rm P}$/MX$_{\rm P}$ for the parallel stacking $\Gamma$ states, otherwise they form a triangular lattice. What remains is to find the magnitude of the moiré-scale onsite repulsion $U^{vs}$.


Commonly, the effective Hubbard $U^{vs}$ relevant for the moir\'e flat bands is derived from the continuum model by constructing a set of continuous Wannier orbitals $w^{vs}_i({\bf r})$ centered at the tight-binding lattice sites $i$, depending on valley $v$ and stacking $s$. Their self-interaction is then $U^{vs}_{ii'} = \int d^2r d^2r' |w^{vs}_i({\bf r})|^2 V({\bf r}-{\bf r}') |w^{vs}_{i'}({\bf r}')|^2$ where $V({\bf r}) = \frac{e^2}{\epsilon} \left[ r^{-1} - (r^2 + D^2)^{-1/2} \right]$ is a gate-screened Coulomb interaction with distance to the gate $D$~\cite{Wu.2018}. However, this neglects all details about the internal screening channels (which enter here only effectively via $\epsilon$ and $D$) and the fact that on the atomic scale the electronic states are not smooth, but in fact are superpositions of $d$-orbitals $\varphi_{j, \alpha} ({\bf r})$ on different W atoms labeled by index $j$. The smooth function $w^{vs}_i({\bf r})$ therefore represents the weights on the different atomic $d$ orbitals, so that the {\em actual} Wannier orbital $\phi^{vs}_i({\bf r})$ becomes
\begin{equation}
    \phi^{vs}_i({\bf r}) = \sum_{j \in \left\{ W \right\}, \alpha} w_i^{vs} (r_j) \varphi_{j, \alpha} ({\bf r} - {\bf r}_j)
\label{Eq:Basis}
\end{equation}
where $j$ sums over all W atoms at positions ${\bf r}_j$, and $\alpha$ labels the relevant atomic orbitals, $\alpha = d_{z^2}$, $d_{x^2 - y^2}, d_{xy}$, etc. In order to estimate the Hubbard $U^{vs}$ for the moiré-scale tight-binding model, we thus need to first calculate the relevant $U_{\alpha \beta}$ for the atomic-scale $d$-orbitals of W atoms.

To this end we performed constrained RPA (cRPA) calculations~\cite{aryasetiawan_frequency-dependent_2004} to extract the non-local Coulomb interactions $U_{j\alpha,j'\beta}(R) = \int d^2r d^2r' |\varphi_{j,\alpha}({\bf r})|^2 U({\bf r}-{\bf r}') |\varphi_{j',\beta}({\bf r}'+R)|^2$ for the six untwisted bilayer stackings of WSe$_2$ in the atomic $d$ orbital basis. This way, we exclude any screening from the $d$ orbitals, but consistently take the screening from all other states into account. In this case, it turns out that the local stacking details, which change throughout the moiré unit cell and with the twisting angle, do not affect the cRPA screened interactions much (see SI~\cite{SupplInfo} for details), similar to the situation in twisted bilayer graphene~\cite{westerhout_quantum_2021}. Using the basis transformation from Eq.~\eqref{Eq:Basis} we can thus derive the effective moiré $U^{vs}_{\rm moir\acute{e}} = \sum_{j,j'} U_{j\alpha,j'\alpha} |w^{vs}(r_j)|^2 |w^{vs}({\bf r}_{j'})|^2$, where $\alpha$ is the relevant $d$-orbital for the given valley and stacking. The twist angle only enters this expression for $U^{vs}$ via the envelope functions $w^{vs}_i({\bf r})$. To this end, we parametrize the onsite cRPA $U_{\alpha\beta}(R)$ with an image-charge model of the form:
\begin{eqnarray}
U( r) &=& \frac{e}{\varepsilon_m}
    \left[
    \frac{1}{\sqrt{r^2+\delta}^2} + \right. \nonumber \\ &&
    \left. 
    2\sum_{n=1}^{\infty}\dfrac{1}{\sqrt{r^2 + {\delta}^2 + (n h)^2}} 
    \Big(\dfrac{{\varepsilon_m}-\varepsilon_\text{env}}{{\varepsilon_m} + \varepsilon_\text{env}}\Big)^n \right],
\label{Eq:FitU}
\end{eqnarray}
where $e$ is the elementary charge, $h$ represents an effective height of the bilayer, $\delta$ is a parameter allowing to fit the atomic local interaction $U_{ii} = U(r = 0)$, $\epsilon_m$ is the effective dielectric constant of the bilayer, and $\epsilon_\text{env}$ represents additional screening from the environment, such as substrates (and all non-Wannierized moiré bands). 
This is similar to the Keldysh-potential model utilized in Ref.~\cite{Ryee.2023}.
In the Supplementary Information~\cite{SupplInfo} we show the results of fitting $h$, $\delta$, and $\varepsilon_m$ to ab initio cRPA calculations for different pressures, stackings, and for different $d$ orbital channels. We find that the fitting parameters and thus the analytic $U( r)$ model is only mildly affected by pressure and stacking, such that we proceed with using a fixed set of fitting parameters, which are only different for the $\Gamma$ and K valley models.
All together, this provides us with a light-weight effective Hubbard model on the moiré scale, with its model parameters depending on pressure, twist angle, valley ($\Gamma$/K) and type of stacking (parallel/antiparallel). 
For the band structures shown in Fig.~\ref{fig:moire} at $\theta = 3^\circ$ twist angle, the relevant interaction parameters for $\epsilon_\text{env} = 30$ and $P=0$ are $U^{{\rm K}, {\rm P}} = 51$ meV, $U^{\Gamma, {\rm P}} = 170$ meV, and $U^{\Gamma, {\rm AP}} = 132$ meV.

In order to estimate the effect of the Coulomb interactions we performed self-consistent Hartree-Fock mean field theory~\cite{SupplInfo}. As discussed in Sec.~\ref{Sec:Moire}, the relevant flat bands in antiparallel stackings as well as in the K-valley for parallel stackings form a {\em triangular} lattice. When $U/t$ is large in the triangular lattice Hubbard model at half-filling, the system becomes a three-sublattice 120$^\circ$ antiferromagnet, which can be qualitatively correctly described by Hartree-Fock theory~\cite{Szasz.2020}. Similarly, the effective {\em honeycomb} lattice Hubbard model relevant for the parallel stacking $\Gamma$ valley exhibits Néel order for large $U/t$ at half-filling. Note, however, that in terms of doping {\em per unit cell}, half-filling for a triangular lattice corresponds to $\nu = 1$ whereas half-filling for the honeycomb lattice occurs at $\nu = 2$. Away from half-filling, ferromagnetism is competing with antiferromagnetism~\cite{Hirsch.1985}.

The resulting phase diagrams for parallel and antiparallel stacked moiré bilayers at $\theta = 3^\circ$ degree twist angle in a dielectric environement with $\epsilon_\text{env} = 30$ are shown in Fig.~\ref{fig:phasediagram}. Both diagrams display a prominent valley charge-transfer under the application of pressure. At low hole density ($\nu <1$ for antiparallel stacking, $\nu < 2$ for parallel stacking) there is a first order transition from states in the K valley to states in the $\Gamma$ valley.

At half-filling of the $\Gamma$ valley, we find a pressure-induced transition into an antiferromagnetic charge-transfer insulator. Here the flat bands in the $\Gamma$ valley are split into the mean-field precursors of lower and upper Hubbard bands. The states of the K valley are now {\em within the gap} of the $\Gamma$ valley states. This is exactly the situation described in the Zaanen-Sawatzky-Allen (ZSA) scheme~\cite{Zaanen.1985}, originally intended for cuprates where the oxygen $p$-orbitals are in the gap of the copper $d$-orbitals' lower and upper Hubbard band, which is also referred to as charge-transfer insulating state. When the pressure is further increased, the K states move to lower energy and out of the $\Gamma$ gap, leading to a regular Mott-Hubbard insulator of the $\Gamma$ states.

Upon doping the charge-transfer insulator state, the additional charge carriers occupy the K valley. The resulting situation resembles that of the Kondo lattice model~\cite{Coleman.2015p0w,Rademaker.2016vc9}, where weakly correlated conduction electrons (from the K valley) coexist with correlated localized electrons (from the $\Gamma$ valley). In our approximate mean field theory picture, the density of states as presented in Fig.~\ref{fig:mft_dos} clearly shows the coexistence of localized and conduction electrons. In the case of the parallel stacking, we also find weak ferromagnetic order in the K valley.

The mean field phase diagram consists thus of a wealth of different correlated phases: at half-filling both types of Zaanen-Sawatzky-Allen insulators, and away from half-filling either doped Mott insulator or Kondo lattice systems. These correlated metallic phases are ill-understood, and likely show a richer behavior than can be inferred from naive mean field theory. However, by identifying where twisted bilayer WSe$_2$ hosts these different correlated conducting phases we pave the way for further {\em quantitative} studies that compare theories and experiments.


\section{Outlook}

In summary, we have found that the dominant structural parameter of bilayer WSe$_2$ is the interlayer distance $d$, as it affects the relative energy of the $\Gamma$ and K valley states. Consequently, in twisted bilayer WSe$_2$ the effective strength of the interaction -- which is much higher in the $\Gamma$ valley than in the K valley -- can be tuned through uniaxial pressure. The resulting valley charge-transfer allows for a multitude of interesting correlated phases.

Recently, the idea of using pressure to induce a valley shift in (twisted) bilayer TMDs has been addressed in two other theoretical publications~\cite{Gao.2023,Olin.2023} and their ab initio results are consistent with our findings in Sec.~\ref{Sec:AbInitio}. However, they do not discuss the strength nor the effect of many-body Coulomb interactions. At the other extreme, a recent work~\cite{Morales-Durán.2023m1} suggesting a fractional Chern insulator in pressurized twisted bilayer WSe2 ignores the valley degree of freedom. Our results bridge this gap and presents the whole picture as we have explored the full interplay between valley charge transfer and correlation effects.

Note that the valley degree of freedom can also be tuned using a perpendicular electric field~\cite{Huang.2016,Ramasubramaniam.2011}. The states at the ${\Gamma}$ valley are more sensitive than K valley states to an applied electric field, as they correspond to orbitals that are more extended in the $z$-direction. However, we find tuning by electric field to be less feasible for a twisted bilayer, as a perpendicular electric field also couples to the layer degree of freedom.


Experimentally, applying pressure on moiré bilayers is shown to be feasible using diamond anvil cells (DACs). This initially started with pressure-studies in twisted bilayer graphene~\cite{Yankowitz.2019} and twisted double bilayer graphene~\cite{Szentpéteri.2021}.
Recently, a hydrostatic pressure experiment was performed on twisted bilayer WSe$_2$~\cite{Xie.2023}. Using exciton photoluminescence~\cite{Mak.2018}, they showed that indeed the energy of the $\Gamma$ valley was shifted upwards, which is consistent with our predictions. A natural next step would be to perform transport experiments in the spirit of the Ref.~\cite{Wang.2020}, revealing new correlated insulator states under pressure.

Among our predicted phases is a Kondo lattice regime with localized electrons in the $\Gamma$ valley and conduction electrons in the K valley. Similar Kondo lattice physics, but in terms of the {\em layer} degree of freedom (instead of the {\em valley} suggested in our work), has been observed in twisted WSe$_2$ homobilayers~\cite{Xu.2022i5} and MoTe$_2$/WSe$_2$ heterobilayers~\cite{Zhao.2023}. The properties of the Kondo lattice many-body phase will strongly depend on the interactions and possible hybridization between the localized and conduction states -- as such it would be interesting to quantitatively compare the layer-Kondo lattice to the valley-Kondo lattice regime, and to traditional heavy fermion systems~\cite{Si.2010,Rademaker.2016vc9} and the `topological' heavy fermions proposed for twisted bilayer graphene~\cite{Datta.2023,Rai.2023}.
Such a comparison, from a theoretical side, would require to go beyond the simple mean-field theory picture presented here. 
Strong correlation methods are needed, such as Dynamical Mean Field Theory~\cite{Georges.1996}, which recently has been used to study TMD moiré bilayers~\cite{Ryee.2023,Tscheppe.2024} as well as other flat band materials~\cite{Crippa.2024,Datta.2023,Rai.2023}.
In addition, a more precise modeling of the inter-valley couplings is required, as well as possible emerging interactions such as the RKKY coupling between the localized electrons. 
Another aspect that pushes beyond our current model is the appearance of strong longer-ranged Coulomb interactions in the $\Gamma$ valley, that can stabilize fractionalized~\cite{Cai.20235nb} or generalized Wigner crystal phases~\cite{Regan.2020,Tan.2023}. 
In any case, such theoretical explorations should be fueled by new experiments on twisted bilayer WSe$_2$ under pressure.


\begin{acknowledgments}
We acknowledge discussions with Abhay Pasupathy and the late Jan Zaanen.
L.R. was funded by the SNSF via Starting grant TMSGI2\_211296.  M.R. acknowledges support from the Dutch Research Council (NWO) via the “TOPCORE” consortium and from the research program “Materials for the Quantum Age” (QuMat) for ﬁnancial support. The latter program (registration number 024.005.006) is part of the Gravitation program ﬁnanced by the Dutch Ministry of Education, Culture and Science (OCW). M.G.\ acknowledges financial support from the Italian Ministry for University and Research through the PNRR project ECS\_00000033\_ECOSISTER and the PRIN2022 project ``Simultaneous electrical control of spin and valley polarization in van der Waals magnetic materials" (SECSY).
The computations were performed at the Swiss National Supercomputing Centre (CSCS) under Projects No. s1146 and at the Dutch National Supercomputer Snellius under Project No. EINF-4184.
\end{acknowledgments}


%

\end{document}